\documentstyle[12pt]{article}
\newcommand{\WTb}{\hspace{5mm}\begin{picture}(36,16)(5,3)\put(3,5)
    {\circle*{4}}\put(23,5)
    {\circle*{4}}\put(3,5){\vector(1,0){10}}\put(13,5){\line(1,0){10}}
    \end{picture}}
\newcommand{\WTa}{\hspace{5mm}\begin{picture}(36,16)(5,3)\put(3,5){\circle*{4}}
   \put(23,5){\circle*{4}}\put(3,5){\line(1,0){10}}\put(23,5){\vector(-1,0){10}}
    \end{picture}}
\newcommand{\WTgI}{\hspace{5mm}\begin{picture}(36,16)(5,3)\put(3,5){\circle*{4}}
    \put(23,5){\circle*{4}}\put(3,1){\circle{4}}
    \end{picture}}
\newcommand{\WTgII}{\hspace{5mm}\begin{picture}(36,16)(5,3)\put(3,5){\circle*{4}}
    \put(23,5){\circle*{4}}\put(23,1){\circle{4}}
    \end{picture}}
\newcommand{\WTbb}{\hspace{5mm}\begin{picture}(36,16)(5,3)\put(3,5){\circle*{4}}
    \put(23,5){\circle*{4}}\put(3,7){\vector(1,0){10}}\put(13,7){\line(1,0){10}}
    \put(3,3){\vector(1,0){10}}\put(13,3){\line(1,0){10}}
    \end{picture}}
\newcommand{\WTaa}{\hspace{5mm}\begin{picture}(36,16)(5,3)\put(3,5){\circle*{4}}
    \put(23,5){\circle*{4}}\put(3,7){\line(1,0){10}}\put(23,7){\vector(-1,0)
    {10}}\put(3,3){\line(1,0){10}}\put(23,3){\vector(-1,0){10}}
    \end{picture}}
\newcommand{\WTab}{\hspace{5mm}\begin{picture}(36,16)(5,3)\put(3,5){\circle*{4}}
    \put(23,5){\circle*{4}}\put(3,7){\vector(1,0){10}}\put(13,7){\line(1,0){10}}
    \put(3,3){\line(1,0){10}}\put(23,3){\vector(-1,0){10}}
    \end{picture}}
\newcommand{\WTggI}{\hspace{5mm}\begin{picture}(36,16)(5,3)\put(3,5){\circle*{4}} 
    \put(23,5){\circle*{4}}\put(3,1){\circle{4}}\put(3,9){\circle{4}}
    \end{picture}}
\newcommand{\WTggII}{\hspace{5mm}\begin{picture}(36,16)(5,3)\put(3,5){\circle*{4}} 
    \put(23,5){\circle*{4}}\put(23,1){\circle{4}}\put(23,9){\circle{4}}
    \end{picture}}
\newcommand{\WTggIxII}{\hspace{5mm}\begin{picture}(36,16)(5,3)\put(3,5){\circle*{4}} 
    \put(23,5){\circle*{4}}\put(3,1){\circle{4}}\put(23,1){\circle{4}}
    \end{picture}}
\newcommand{\WTbgI}{\hspace{5mm}\begin{picture}(36,16)(5,3)\put(3,5){\circle*{4}} 
    \put(23,5){\circle*{4}}\put(3,1){\circle{4}}\put(3,5){\vector(1,0){10}}
    \put(23,5){\line(-1,0){10}}
    \end{picture}}
\newcommand{\WTbgII}{\hspace{5mm}\begin{picture}(36,16)(5,3)\put(3,5){\circle*{4}} 
    \put(23,5){\circle*{4}}\put(23,1){\circle{4}}\put(3,5){\vector(1,0){10}}
    \put(23,5){\line(-1,0){10}}
    \end{picture}}
\newcommand{\WTagI}{\hspace{5mm}\begin{picture}(36,16)(5,3)\put(3,5){\circle*{4}} 
    \put(23,5){\circle*{4}}\put(3,1){\circle{4}}\put(3,5){\line(1,0){10}}
    \put(23,5){\vector(-1,0){10}}
    \end{picture}}
\newcommand{\WTagII}{\hspace{5mm}\begin{picture}(36,16)(5,3)\put(3,5){\circle*{4}} 
    \put(23,5){\circle*{4}}\put(23,1){\circle{4}}\put(3,5){\line(1,0){10}}
    \put(23,5){\vector(-1,0){10}}
    \end{picture}}
\newcommand{\WTbbb}{\hspace{5mm}\begin{picture}(36,16)(5,3)\put(3,5){\circle*{4}}
    \put(23,5){\circle*{4}}\put(3,7){\vector(1,0){10}}\put(23,7)
    {\line(-1,0){10}}\put(3,5){\vector(1,0){10}}\put(23,5){\line(-1,0){10}}
    \put(3,3){\vector(1,0){10}}\put(23,3){\line(-1,0){10}}
    \end{picture}}
\newcommand{\WTaaa}{\hspace{5mm}\begin{picture}(36,16)(5,3)\put(3,5){\circle*{4}}
    \put(23,5){\circle*{4}}\put(3,7){\line(1,0){10}}\put(23,7)
    {\vector(-1,0){10}}\put(3,5){\line(1,0){10}}\put(23,5){\vector(-1,0){10}}
    \put(3,3){\line(1,0){10}}\put(23,3){\vector(-1,0){10}}
    \end{picture}}
\newcommand{\WTaab}{\hspace{5mm}\begin{picture}(36,16)(5,3)\put(3,5){\circle*{4}}
    \put(23,5){\circle*{4}}\put(3,7){\line(1,0){10}}\put(23,7)
    {\vector(-1,0){10}}\put(3,5){\line(1,0){10}}\put(23,5){\vector(-1,0){10}}
    \put(3,3){\line(1,0){10}}\put(23,3){\vector(-1,0){10}}
    \end{picture}}
\newcommand{\WTabb}{\hspace{5mm}\begin{picture}(36,16)(5,3)\put(3,5){\circle*{4}}
    \put(23,5){\circle*{4}}\put(3,7){\vector(1,0){10}}\put(23,7)
    {\line(-1,0){10}}\put(3,5){\vector(1,0){10}}\put(23,5){\line(-1,0){10}}
    \put(3,3){\line(1,0){10}}\put(23,3){\vector(-1,0){10}}
    \end{picture}}
\newcommand{\WTgggI}{\hspace{5mm}\begin{picture}(36,16)(5,3)\put(3,5){\circle*{4}}
    \put(23,5){\circle*{4}}\put(3,1){\circle{4}}\put(3,9){\circle{4}}
    \put(-1,5){\circle{4}}
    \end{picture}}
\newcommand{\WTgggII}{\hspace{5mm}\begin{picture}(36,16)(5,3)\put(3,5){\circle*{4}}
    \put(23,5){\circle*{4}}\put(23,1){\circle{4}}\put(23,9){\circle{4}}
    \put(27,5){\circle{4}}
    \end{picture}}
\newcommand{\WTgggIIxI}{\hspace{5mm}\begin{picture}(36,16)(5,3)\put(3,5){\circle*{4}}
    \put(23,5){\circle*{4}}\put(3,1){\circle{4}}\put(3,9){\circle{4}}
    \put(23,1){\circle{4}}
    \end{picture}}
\newcommand{\WTgggIxII}{\hspace{5mm}\begin{picture}(36,16)(5,3)\put(3,5){\circle*{4}}
    \put(23,5){\circle*{4}}\put(3,1){\circle{4}}\put(23,1){\circle{4}}
    \put(23,9){\circle{4}}
    \end{picture}}
\newcommand{\WTbbgI}{\hspace{5mm}\begin{picture}(36,16)(5,3)\put(3,5){\circle*{4}}
    \put(23,5){\circle*{4}}\put(3,7){\vector(1,0){10}}\put(23,7)
    {\line(-1,0){10}}\put(3,3){\vector(1,0){10}}\put(23,3){\line(-1,0){10}}
    \put(3,1){\circle{4}}
    \end{picture}}
\newcommand{\WTbbgII}{\hspace{5mm}\begin{picture}(36,16)(5,3)\put(3,5){\circle*{4}}
    \put(23,5){\circle*{4}}\put(3,7){\vector(1,0){10}}\put(23,7)
    {\line(-1,0){10}}\put(3,3){\vector(1,0){10}}\put(23,3){\line(-1,0){10}}
    \put(23,1){\circle{4}}
    \end{picture}}

\newcommand{\WTbggI}{\hspace{5mm}\begin{picture}(36,16)(5,3)\put(3,5){\circle*{4}}
    \put(23,5){\circle*{4}}\put(3,5){\line(1,0){10}}\put(23,5)
    {\vector(-1,0){10}}\put(3,1){\circle{4}}\put(3,9){\circle{4}}
    \end{picture}}
\newcommand{\WTbggII}{\hspace{5mm}\begin{picture}(36,16)(5,3)\put(3,5){\circle*{4}}
    \put(23,5){\circle*{4}}\put(3,5){\line(1,0){10}}\put(23,5)
    {\vector(-1,0){10}}\put(23,1){\circle{4}}\put(23,9){\circle{4}}
    \end{picture}}
\newcommand{\WTbggIxII}{\hspace{5mm}\begin{picture}(36,16)(5,3)\put(3,5){\circle*{4}}
    \put(23,5){\circle*{4}}\put(3,5){\line(1,0){10}}\put(23,5)
    {\vector(-1,0){10}}\put(3,1){\circle{4}}\put(23,1){\circle{4}}
    \end{picture}}
\newcommand{\WTaggI}{\hspace{5mm}\begin{picture}(36,16)(5,3)\put(3,5){\circle*{4}}
    \put(23,5){\circle*{4}}\put(3,5){\line(1,0){10}}\put(23,5)
    {\vector(-1,0){10}}\put(3,1){\circle{4}}\put(3,9){\circle{4}}
    \end{picture}}
\newcommand{\WTaggII}{\hspace{5mm}\begin{picture}(36,16)(5,3)\put(3,5){\circle*{4}}
    \put(23,5){\circle*{4}}\put(3,5){\line(1,0){10}}\put(23,5)
    {\vector(-1,0){10}}\put(23,1){\circle{4}}\put(23,9){\circle{4}}
    \end{picture}}
\newcommand{\WTaggIxII}{\hspace{5mm}\begin{picture}(36,16)(5,3)\put(3,5){\circle*{4}}
    \put(23,5){\circle*{4}}\put(3,5){\line(1,0){10}}\put(23,5)
    {\vector(-1,0){10}}\put(3,1){\circle{4}}\put(23,1){\circle{4}}
    \end{picture}}

\begin{document}
\title{Time Machines and the Breakdown of Unitarity}
\author{Frank Antonsen and Karsten Bormann\thanks{presently at
the Technical University of Denmark}\\ 
The Niels Bohr Institute\\ Blegdamsvej 17, DK-2100 Copenhagen \O,
Denmark}
\date{}
\maketitle
\newpage

\begin{abstract}
We present a generic way of thinking about time machines from the view
of a far away observer. In this model
the universe consists of three (or more) regions: One containing the
entrance of the time machine, another the exit and the remaining one(s)
the rest of the universe. In the latter we know ordinary quantum mechanics
to be valid and thus are able to write down a Hamiltonian describing this 
generic time machine. We prove the time-evolution operator to be non-symmetric.
Various interpretations of this irreversibility are given.
\end{abstract}

\section*{Introduction}
The question of whether time machines are possible or not has 
been studied by several authors in the last couple of years. 
This interest was spawned
by the realization that topologically non-trivial
space-times may exhibit closed time-like curves, or ``time machines''.
The most important example is an otherwise flat spacetime
with a sufficiently short wormhole connecting two distant regions, 
see figure 1. 
This can be made to function as a
time-machine, either by putting the two mouths of the
wormhole in regions of different gravitational potential
or by accelerating one with respect to the other, and then
bringing it to rest again -- both of these methods generate
a time shift, which an object travelling through the wormhole
experiences (Morris {\em et al.} (1988), Kim and Thorne (1991),
Friedman {\em et al.} (1990), Novikov (1992)). 
\\
The presence of closed time-like curves (time-machines)
would make the past and the future fuse in the sense
that `someone' travelling on a closed time-like curve
could influence his own past (the past and future light cones
overlap). So time-machines makes
distinguishing past and future impossible, right?
Wrong! The
Hamiltonian describing the action of the time-machine
becomes non-symmetric making the evolution operator non-unitary, and thus
time machines will be time-asymmetric in a quantum mechanical context.
\\
We can model such time machines very easily. First assume a 3+1 splitting
of spacetime, i.e. the existence of a cosmic time (if the time machine is
constructed from a wormhole, then this splitting will only be possible
sufficiently far away from the mouths). Space will be divided 
into a number of regions. The time machine has its entrance (deep) inside region
1, its exit (deep) within region 2, see figure 2, and it operates in the
following way: any object entering a particular region, region 1, 
at time $t$, reappears in another region, 2, with a probability $\alpha$
but at time $t-T$, i.e. it has moved backwards in time. 
Similarly, an object entering region 2 at a time $t$
will reappear in region 1 at time $t+T$ with a probability $\beta$, 
i.e. it has moved forward in time. This is the essence of what a 
time machine does, and is the only effect we are going to study in this paper.
These two regions, 1 and 2, could contain the mouths of a wormhole, and 
we will often refer to them as the ``mouths'' of the time machine. 
No assumption is made concerning 
the actual structure of the time machine, it could be a wormhole or it could be
something else. \\
The objects will be taken to be the quanta
of some scalar field (one could with very little extra trouble  -- or gain --
treat quanta of arbitrary spin too). It will be shown
that the number of particles entering the wormhole
is different from the number coming out in the other
end which is most unfortunate. Thus time machines 
make it possible to distinguish past and 
future, by for instance looking at the density of 
some Bose field initially distributed homogeneously in space. 
They also pose a threat to energy conservation. Of course one could put this
difference in particles/energy into the time machine's
internal structure in order to have energy conservation
- getting the extra particles out/in would thus be classifiable as part of the 
maintenance costs, but to an external observer
a neglected time machine looks like an energy source/drain.
If they are homogeneously distributed, this observation makes the existence of 
wormholes (or any other structure capable of supporting a
time machine) with sizes in the interval between
$\sim 10^{-18}$ m and $\sim 10^{8}$ m highly unlikely
--- they would have been observed. It also  
makes it dubious whether a ``time machine'' would
really be up to its name, i.e. whether a space-time
possessing closed time-like curves, would function
as a time machine in the traditional sense of the word.

\section*{Breakdown of Unitarity in the Presence
of Time Machines}
We consider a partition of space, and we label each of these regions such
that region 1 is one of the ``mouths'' and region 2 the other. We assume that
particles entering region 1 will reappear, with some probability, in region 
2 but at an earlier time and vice versa. Since the ``mouths'' are assumed
to lie deep within the appropriate regions, these probabilities, $\alpha,
\beta$ will typically not be one, i.e. $\alpha,\beta <1$. The time step
will be assumed identical in both directions and will be denoted by $T$. This
is not a severe assumption: if the time steps were different in the two 
directions, non-unitarity would be obvious. The
Hamiltonian will be taken to be the simplest possible, namely a slight
generalisation of the canonical Hamiltonian of a free field in number 
representation:
\begin{equation}
    H=\alpha a_1^\dagger(t+T)a_2(t)+\beta a_2^\dagger(t-T)a_1(t) + 
	g\sum_{i=1}^Na_i^\dagger(t)a_i(t)
\end{equation}
with $i$ labelling the various regions, $i=1,2,...,N$ where $N$ could be 
infinite (it has to be at least three: the two ``mouths'' and the rest of 
the universe). 
Here the $g$-term simply counts the number of quanta
in the various regions at time $t$, whereas the $\alpha,\beta$-terms 
describe the actual time machine effect.\\
Had $i$ been the momentum and 
had $t$ instead referred to a particular site in a chain, then
this would be a familiar Hamiltonian -- the first two terms would be 
``hopping terms'' 
describing the possibility of a quanta to jump from one site to another. \\
We consider the regions 1 and 2 as
identified modulo a time-shift, which implies the following commutator 
relations (assuming bosonic statistics)
\begin{equation}
    \left[a_i(t),a_j^\dagger(t')\right] = \delta_{ij}\Delta(t-t') 
      +\delta_{i1}\delta_{j2}
    \Delta(t'-t+T)+\delta_{i2}\delta_{j1}\Delta(t'-t-T)\qquad
      i,j = 1,2,...,N
\end{equation}
the remaining commutators all vanishing.\footnote{Thus 
the time machine gives rise to two modifications,
(1) the presence of the $\alpha,\beta$-terms in the Hamiltonian, 
and (2) the $\Delta(t'-t\pm T)$-terms in the commutator relations. 
These two modifications are of course not independent: putting
either $\alpha$ or $\beta$ equal to zero amounts to forbidding travel 
through the wormhole in the corresponding direction, and hence the analogous 
term in the commutator relations should also
be removed. To avoid a too heavy notation, we have decided, however, 
not to let this appear explicitly in equations (1) and (2).} 
The function $\Delta$ is a (possibly) smeared
Dirac delta-distribution, the smearing mimicking some uncertainty in 
the values of $t,t'$. Its precise form matters little for our calculation;
it could just as well be a proper Dirac
delta-distribution. Our lack of knowledge about the precise structure of the 
time-machine can be parametrised by this function $\Delta(t)$ and the 
coefficients $\alpha,\beta$ appearing in the Hamiltonian.
So the second quantisation operators corresponding to different regions 
at different time commute, except for those corresponding to the ``mouths''.\\
The time evolution operator $U(t,t')$
is given by $U(t,t')=U(t-t') = e^{-iH(t-t')}$ and hence we need to evaluate
powers of $H$. We want to find the matrix elements of $U(t,t')$.
Denoting the states by $|n,t\rangle$,
with $n=(n_1,n_2,...,n_N)$ a multi index describing the number of 
quanta in each region, we have
\begin{eqnarray}
    \langle n,t|H|n',t'\rangle &=& \alpha\delta_{n_2',n_2-1}\delta_{n_1',n_1+1}
    \Delta(t'-t-T)\sqrt{n_2(n_1+1)}\prod_{i\neq 1,2}\delta_{n_i',n_i}+\nonumber
    \\
    &&\beta \delta_{n_2',n_2+1}\delta_{n_1',n_1-1}\Delta(t'-t+T)\sqrt{n_1n_2}
    \prod_{i\neq 1,2}\delta_{n_i',n_i}+\nonumber\\
    &&g\Delta(t-t')\delta(n,n')\sum_i n_i
\end{eqnarray}
where $\delta(n,n')\equiv\prod_i\delta_{n_i,n_i'}$ is a Kronecker delta.\\  
Similarly we get
\begin{eqnarray}
    \langle n,t|H^2|n',t'\rangle &=& \alpha^2\delta_{n_1',n_1-2}\delta_{n_2',
    n_2+2}\sqrt{(n_1-2)(n_1-3)(n_2+1)(n_2+2)}\Delta(t'-t+T)\delta_{12}(n,n')
    \nonumber\\
    &&+\alpha^2\delta_{n_1',n_1-1}\delta_{n_2',n_2+1}\sqrt{(n_1-1)(n_2+1)}
    \Delta(t'-t+T)\delta_{12}(n,n')\nonumber\\
    &&+\beta^2\delta_{n_1',n_1+2}\delta_{n_2',n_2-2}\sqrt{(n_1+3)(n_1+4)n_2
    (n_2-1)}\Delta(t'-t-T)\delta_{12}(n,n')\nonumber\\
    &&+\beta^2\delta_{n_1',n_1+1}\delta_{n_2',n_2-1}\sqrt{(n_1-1)(n_2+1)}
    \Delta(t'-t-T)\delta_{12}(n,n')\nonumber\\
    &&+g^2\delta(n,n')\Delta(t-t')\sum_{i\neq j}n_in_j + g^2\delta(n,n')
    \Delta(t-t')\sum_i n_i(n_i+1)\nonumber\\
    &&+\alpha\beta\left(n_2(n_1+1)+n_1n_2\right)\delta_{12}(n,n')\delta
    (|t-t'|-T)\nonumber\\
    &&+\alpha g\delta_{12}(n,n')\sum_in_i\left(\delta_{n_1',n_1-1}
    \delta_{n_2',n_2+1}\Delta(t'-t+T)\sqrt{(n_1-1)(n_2+1)}+1\right)\nonumber\\
    &&+\beta g\sum_in_i\left(\delta_{n_1',n_1+1}\delta_{n_2',n_2-1}
    \Delta(t'-t-T)\sqrt{(n_2-1)(n_1+1)}+1\right)
\end{eqnarray}
with$\delta_{12}(n,n') \equiv \prod_{i\neq 1,2}\delta_{n_i',n_i}$.
The time asymmetry of the Hamiltonian thus manifests itself in the evolution
operator. This will be seen even more clearly in the next order contribution.\\
 A convenient way of representing the
various contributions are in terms of diagrams as follows: the two regions 
1 and 2 are
represented by two dots, $\bullet$ -- the remaining $N-2$ regions need not 
be drawn, as they 
are not influenced by the time machine -- the particle motion is then 
indicated by
arrows, the $g$-terms counts the number of particles and are essentially vacuum
terms, they are represented by closed loops. This gives the diagrams listed in
table 1. We refer to these as ``worm tracks'' (again thinking of the time 
machine
as being made from a wormhole). Table 2 shows the various
contributions to $H^3$ (here $t_\pm\equiv t\pm T$). We see that we generate
asymmetries even in these {\em low order} terms. The worm tracks and the 
weights with which they appear are listed in table 3. \\
The Hamiltonian itself, is of course not a symmetric operator, as it 
identifies two
different regions provided there is specific difference between the times, but
when calculating the higher powers of $H$ we discover new asymmetries, which
were not to be expected {\em a priori}. This is so even in the most symmetric
case $\beta=\alpha$, in fact the result is quite independent of what
the precise values of the parameters $\alpha, \beta, g$ are.\\
It follows from eqs(3,4) and tables 2-3 that more quanta are exiting the
time machine than there are entering it.
The non-symmetric nature of the Hamiltonian thus generates, through the time
evolution operator, an irreversibility, which is surprisingly strong. \\

\subsection*{Generation of Entropy}
Non-symmetric time evolution is usually taken to be a sign of irreversibility 
and hence of entropy generation. We want to show that this is certainly so
in our case, at least to the very lowest order.\\
Given a density matrix, $\rho$, the entropy is
\begin{equation}
	S = -{\rm Tr}~ \rho\ln\rho
\end{equation}
In our case $\rho$ is (up to a normalisation constant) just the time evolution
operator $U(t,t')$. Thus we can use our expressions for the matrix
elements of the Hamiltonian found above. First of all we notice that the terms
only involving the $g$-contributions correspond to a free field configuration
and consequently have vanishing entropy change (if all the contributions are
added together). We only need to concentrate on the contributions involving 
the $\alpha,\beta$-part. This is also what one would expect, as these are 
precisely the time machine specific parts of the Hamiltonian. Furthermore,
since a trace is involved in the definition of $S$ we only need to keep the
diagonal parts of $\langle n,t|H^k|n',t'\rangle$. Since $\rho\ln\rho \sim
UH$, the first such contribution comes from the matrix elements of $H^2$.
Thus
\begin{eqnarray}
	{\rm Tr} UH &=& g\Delta(t-t')\sum_n n - (t-t')\left(g^2\Delta(t-t')
	\sum_{n_i,n_j}n_in_j+\right.\nonumber\\
	&&\qquad\left.\alpha\beta\Delta(|t-t'|-T)\sum_{n_1,n_2}
	(n_2(n_1+1)+n_1n_2)+...\right)
\end{eqnarray}
The only surviving term is seen to be (as mentioned above, the $g$-terms will
vanish when one takes all powers into account)
\begin{equation}
	(t-t') \alpha\beta \Delta(|t-t'|-T)\sum_{n_1,n_2}(n_2(n_1+1) + n_1n_2)
\end{equation}
Now, this sum is divergent and need to be regularized. The obvious 
regularization scheme to choose is $\zeta$-function regularization, (Hawking 
(1977), Ramond (1989)). One replaces sums like
\begin{displaymath}
	\sum_n n^{-s}
\end{displaymath}
by a Riemann $\zeta$-function, $\zeta(s)$. This can be analytically continued
to values of $s$ where the above, unregularized summation is illdefined.\\
In our case we need
\begin{equation}
	\left(\sum_n n \right)_{\rm reg} = \zeta(-1) = -\frac{1}{12}
\end{equation}
and similarly
\begin{equation}
	\left(\sum_n(n+1)^{-s}\right)_{\rm reg} = \zeta(s,1)
\end{equation}
where $\zeta(s,a)$ is the so-called Hurwitz $\zeta$-function. We only need
to know the value at $a=1,s=-1$, corresponding to a regularized value for
$\sum_{n_1}(n_1+1) :=\zeta(-1,1)=-\frac{1}{12}$. Hence the regularized
contribution to the entropy reads
\begin{equation}
	(t-t') \alpha\beta \Delta(|t-t'|-T)\sum_{n_1,n_2}(n_2(n_1+1) + n_1n_2)
	:= \frac{1}{72}(t-t')\alpha\beta\Delta(|t-t'|-T)
\end{equation}
Whenever $\alpha\beta>0$ this is positive, and hence we have created entropy.\\
Hence, time-machines can generate entropy and will consequently generate an
arrow of time, contrary to what one would expect.

\section*{Time-Evolution of Operators and Generalised Bogulyubov
Transformations}
From the commutator relations it is straightforward to derive the equations of
motion for the operators $a_1,a_1^\dagger,a_2,a_2^\dagger$. These turn out 
to be
\begin{eqnarray}
    i\dot{a}_1(t) &=& -(\beta+g)a_1(t)\\
    i\dot{a}_1^\dagger(t) &=& \beta a_2^\dagger(t-T)+ga_1^\dagger(t)\\
    i\dot{a}_2(t) &=& -(\alpha+g)a_2(t)\\
    i\dot{a}_2^\dagger &=& \alpha a_1^\dagger(t+T)+ga_2^\dagger(t)
\end{eqnarray}
in which the asymmetry is also apparent. We can diagonalise these by means of a
generalised Bogulyubov transformation. Write
\begin{eqnarray}
    b_1^\dagger(t) &=& U_{11}(t)a_1^\dagger(t)+U_{12}(t)a_2^\dagger(t-T)\\
    b_2^\dagger(t) &=& U_{21}(t)a_1^\dagger(t+T)+U_{22}(t)a_2^\dagger(t)
\end{eqnarray}
while the annihilation operators are not transformed. The transformation 
matrix $U(t)$ then has to satisfy
\begin{eqnarray}
    i\frac{d}{dt}\left(\begin{array}{c} U_{11}\\U_{12} \end{array}\right) &=&
    \left(\begin{array}{cc}\omega_1-g & -\alpha\\ -\beta & \omega_1-g
	\end{array}
    \right)\left(\begin{array}{c}U_{11}\\U_{12}\end{array}\right)\\
    i\frac{d}{dt}\left(\begin{array}{c} U_{21}\\U_{22} \end{array}\right) &=&
    \left(\begin{array}{cc}\omega_2-g & -\alpha\\ -\beta & \omega_2-g
	\end{array}
    \right)\left(\begin{array}{c}U_{21}\\U_{22}\end{array}\right)
\end{eqnarray}
where $\omega_1,\omega_2$ are the energies. Solving these equations is an easy
matter (the coefficients $\omega_1,\omega_2,\alpha,\beta,g$ are all 
constants). The new operators then satisfy
$\dot{b}_i^\dagger = -i\omega_ib_i^\dagger,~i=1,2$.\\
Thus, considering the four operators $a_i,a_i^\dagger$ as independent, we can
make a transformation, unto ``normal modes'', $b_i^\dagger, a_i$,
the energies of which are
$\omega_1, \omega_2, -(\alpha+g),-(\beta+g)$. This means that we {\em can} 
transform the
Hamiltonian unto a diagonal form, using a kind of generalised normal modes,
but these modes will manifestly break hermiticity, as then $(a_i)^\dagger =
a_i^\dagger
\neq b_i^\dagger$ -- the quanta annihilated by $a_i$ are not the same as those
created by $b_i^\dagger$. This is also seen in the fact that the 
``energies'' of the
operators $b_i^\dagger$ (i.e. $\omega_1,\omega_2$) need not be identical to
that of the $a_i$ (i.e. $-(\alpha+g),-(\beta+g)$). Since ``switching off''
the time machine forces $(a_i)^\dagger = b_i^\dagger$ and the energies to be
identical, the time-machine is then seen as a mechanism that forces
$b_i^\dagger$ away from $(a_i)^\dagger$ for $i=1,2$ (or equivalently as 
driving 
$\omega_i$ away from $-(\alpha+g),-(\beta+g)$), thereby generating
non-unitarity of the time-evolution operator. We note that in this 
``diagonalised'' representation of the Hamiltonian, the explicit reference to
the time-shift $T$ has disappeared; it will only enter if one transforms back
to the original basis. 

\section*{Conclusion}
We assumed the existence of some kind of cosmic time (the 3+1 splitting) at 
least sufficiently far away from regions 1 and 2. But this cosmic time will
{\em a priori} not have a particular direction -- both the laws of relativity
and of quantum mechanics are invariant under time-reflections.
It is therefore rather surprising that the presence of time
machines, that above all is seen as destroying causality,
creates an irreversibility and thus, to be consistent with the second law of
thermodynamics, imposes the arrow of time.
\\
However, this is not the only physical effect of
such time machines.  Also basic
subjects of physics are influenced on top of the
problems with causality. Notably, in quantum field
theory unitarity is broken (this is actually
due to the breakdown of causality) and renormalisation
theory will need a modification due to the 
emergence of topologically in-equivalent loop-diagrams,
some of which it is not {\em a priori} possible to do 
away with as they stem from the breakdown of
causality.\\
There is also a problem with the conservation of energy. Since more quanta
are leaving than entering the time machine regions, energy has to supplied
in order to have energy conservation. This need for constantly supplying 
energy will, quite irrespective of the problems of actually avoiding the
energy from traversing the time machine, thus exacerbates the maintenance 
cost making them even more unstable than previously thought
(Antonsen, Bormann (1995 and 1996)).\\
We emphasise that these conclusions are quite generic as any time machine will,
from a bird's eye view, behave as the model presented here.

\subsection*{References}
F. Antonsen, K. Bormann (1995): Int.J. Theor.Phys. {\bf 34} (1995) 2061.\\
F. Antonsen, K. Bormann (1996): Int.J. Theor.Phys. {\bf 35} (1996) 1223.\\
F. Echeverria, G. Klinkhammer and K. Thorne (1991): Phys.Rev. {\bf D44} 
(1991) 1077.\\
J. Friedman {\em et al.} (1990): Phys.Rev.{\bf D42} (1990) 1915.\\
S. Hawking (1977): Commun. Math. Phys. {\bf 56} (1977) 133.\\ 
S.-W. Kim and K.S. Thorne (1991): Phys.Rev. {\bf D43} (1991) 3929.\\
M.S. Morris, K.S. Thorne and U. Yurtsever (1988): Phys.Rev.Lett. {\bf D61} 
(1988) 1447.\\
I.D. Novikov (1992): Phys.Rev. {\bf D45} (1992) 1989.\\
P. Ramond (1989): {\em Field Theory: A Modern Primer/2ed}, Addison-Wesley, 
Redwood City). 

\setlength{\unitlength}{2pt}
\thicklines
\newpage
\pagestyle{empty}
\begin{table}[htb]
\centering
\begin{tabular}{|l|l|l|}\hline
power of $H$ & term & wormtracks\\ \hline
      & $\alpha$ & \WTa\\
$H$   & $\beta$  & \WTb\\
      & $g$      & \WTgI + \WTgII\\ 
      &&\\ \hline
      & $\alpha^2$ & \WTaa + \WTa\\
      & $\beta^2$  & \WTbb + \WTb\\
      & $g^2$      & \WTggI + \WTggII + \WTggIxII\\
$H^2$ & $\alpha\beta$ & \WTab \\
      & $\alpha g$ & \WTagI + \WTagII\\
      & $\beta g$  & \WTbgI + \WTbgII\\ 
      &&\\ \hline
\end{tabular}
\caption{The wormtracks corresponding to the various contributions to $H$ and
$H^2$. The filled out circles represents the regions 1 and 2 respectively, while
open circles represent a number operator and arrows a motion of a particle as
described in the text.}
\end{table}

\newpage
\begin{table}[htb]
\centering
\begin{tabular}{|c|l|}\hline
term & contribution to $H^3$\\ \hline
$\alpha^3$ & $a_1^\dagger(t_+)a_1^\dagger(t_+)a_1^\dagger(t_+)a_2a_2a_2 + 
3a_1^\dagger(t_+)a_1^\dagger(t_+)a_2a_2+a_1^\dagger(t_+)a_2$\\
$\beta^3$ & $a_1a_1a_1a_2^\dagger(t_-)a_2^\dagger(t_-)a_2^\dagger(t_-) - 
6a_1a_1a_2^\dagger(t_-)a_2^\dagger(t_-)+7a_1a_2^\dagger(t_-)$\\
$g^3$ & $\sum_{ijk}a_i^\dagger a_j^\dagger a_k^\dagger a_ia_ja_k + 3\sum_{i,j}
a_i^\dagger a_j^\dagger a_ia_j+\sum_i n_i$\\ \hline
$\alpha^2\beta$ & $3a_1^\dagger(t_+)a_1^\dagger(t_+)a_2a_2a_1a_2^\dagger(t_-) -3
a_1^\dagger(t_+)a_1^\dagger(t_+)a_2a_2+a_1^\dagger(t_+)a_2a_1a_2^\dagger(t_-) 
-a_1^\dagger(t_+)a_2$\\
$\alpha\beta^2$ & $3a_1^\dagger(t_+)a_2a_1a_1a_2^\dagger(t_-)a_2^\dagger(t_-)
-9a_1^\dagger(t_+)a_2a_1a_2^\dagger(t_-)+3a_1^\dagger(t_+)a_2$\\
$\alpha^2g$ & $3\sum_ja_1^\dagger(t_+)a_1^\dagger(t_+)a_j^\dagger a_2a_2a_j + 
3a_1^\dagger(t_+)a_1^\dagger(t_+)a_2a_2+2\sum_ja_1^\dagger(t_+)a_j^\dagger a_j
a_2+$\\
    & $3a_1^\dagger(t_+)a_2^\dagger a_2a_2+a_1^\dagger(t_+)a_2+n_2$\\
$\alpha g^2$ & $3\sum_{jk}a_1^\dagger(t_+)a_j^\dagger a_k^\dagger a_2a_k +
6\sum_ja_1^\dagger(t_+)a_j^\dagger a_2a_j + a_1^\dagger(t_+)a_2 + 3\sum_j
a_2^\dagger a_j^\dagger a_ja_2+2n_2$\\
$\alpha\beta g$ & $8\sum_{j}a_1^\dagger(t_+)a_j^\dagger a_2a_1a_j
a_2^\dagger(t_-)-9a_1^\dagger(t_+)a_2-2a_2^\dagger a_1 -n_2-9a_1^\dagger(t_+)
a_2a_1a_2^\dagger(t_-)-$\\
    & $3a_1^\dagger(t_+)a_1^\dagger a_2a_2 + a_2^\dagger a_2a_2a_2^\dagger(t_-)
-\sum_ja_1^\dagger(t_+)a_j^\dagger a_ja_1 +2a_2^\dagger a_2a_1a_2^\dagger(t_-)
-\sum_ja_1^\dagger a_j^\dagger a_ja_2$\\ \hline
$\beta^2 g$ & $3\sum_j a_j^\dagger a_1a_ja_1a_2^\dagger(t_-)a_2^\dagger(t_-)
-2a_1^\dagger a_1a_2a_2^\dagger(t_-)-9\sum_ja_j^\dagger a_ia_ja_2^\dagger(t_-)
+$\\
    & $3a_aa_1a_2^\dagger(t_-)a_2^\dagger(t_-)-10a_1a_2^\dagger(t_-)
    +4n_1+3\sum_jn_j+3$\\
$\beta g^2$ & $2\sum_{jk}a_j^\dagger a_k^\dagger a_1a_ja_ka_2^\dagger(t_-) -
2\sum_j a_j^\dagger a_1^\dagger a_1a_j+5\sum_ja_j^\dagger a_1a_ja_2^\dagger
(t_-)-$\\ 
    & $\sum_{jk}a_j^\dagger a_k^\dagger a_ja_k-4\sum_jn_j -
    3n_1+a_1a_2^\dagger(t_-)-1$\\ \hline
\end{tabular}
\caption{The contributions to $H^3$. Only shifted times, $t_\pm = t\pm T$, 
are written explicitly. Some of these terms will have vanishing matrix
elements.}
\end{table}

\setlength{\unitlength}{1.5pt}
\begin{table}[htb]
\centering
\begin{tabular}{|l|l|}\hline
term & wormtracks and their weights\\ \hline
$\alpha^3$ & \WTaaa +3$\times$\WTaa + \WTa \\
$\beta^3$  & \WTbbb -6$\times$\WTbb + 7$\times$ \WTb\\
$g^3$      & \WTgggI + \WTgggII + \WTgggIxII + \WTgggIIxI\\
           &+4$\times$\WTggI +4$\times$\WTggII + 4$\times$\WTggIxII\\
           &+ 5$\times$\WTgI +5$\times$ \WTgII \\
           & \\ \hline
$\alpha^2\beta$ & 3$\times$\WTaab -3$\times$\WTaa + \WTab - \WTa\\
$\alpha\beta^2$ & 3$\times$\WTabb - 9$\times$\WTab +3$\times$\WTa\\
$\alpha^2g$     & 3$\times$\WTaggI + 3$\times$\WTaggIxII + 3$\times$\WTaggII
+9$\times$\WTagI\\
    & +9$\times$\WTagII + 10$\times$\WTa + 3$\times$\WTggIxII\\
    &+ 3$\times$\WTggII + 5$\times$\WTgII\\
$\alpha\beta g$ & 8$\times$\WTaggI +8$\times$\WTaggIxII +8$\times$\WTaggII\\
    &+7$\times$\WTagI +7$\times$\WTagII -2$\times$\WTa +2$\times$\WTbgII\\
    &-9$\times$\WTab +3$\times$\WTbb - \WTgI\\
    &-2$\times$\WTgII -\WTggIxII -\WTggI - \WTggII\\
& \\ \hline
$\beta^2 g$ & 3$\times$\WTbbgI +3$\times$\WTbbgII +6$\times$\WTbb\\
    &-9$\times$\WTbgI -9$\times$\WTbgII -19$\times$\WTb\\
    &+7$\times$\WTgI + 3$\times$\WTgII\\
$\beta g^2$ & 2$\times$\WTbggI +2$\times$\WTbggIxII + 2$\times$\WTbggII\\
    &+7$\times$\WTbgI + 7$\times$\WTbgII +8$\times$\WTb\\
    &-3$\times$\WTggI - 3$\times$\WTggIxII -\WTggII\\
    &-10$\times$\WTgI - 5$\times$\WTgII\\
    & \\ \hline
\end{tabular}
\caption{The wormtracks corresponding to the various contributions to $H^3$,
only operator products which involve either region 1 or 2 or both (and 
hence not
terms such as, say, $a_3^\dagger a_4$) are shown. Note the asymmetry.}
\end{table}
\end{document}